# Critical behaviour of the thermal properties of $KMnF_3$

A. Salazar, M. Massot and A. Oleaga

*Departamento de Física Aplicada I, Escuela Técnica Superior de Ingeniería, Universidad del País Vasco, Alameda Urquijo s/n, 48013 Bilbao, Spain*

A. Pawlak

*Institute of Physics, A. Mickiewicz University, Poznan, Poland*

W. Schranz

*Faculty of Physics, University of Vienna, Strudlhofgasse 4, A-1090 Wien, Austria*

E-mail: agustin.salazar@ehu.es

**Abstract**

An ac photopyroelectric calorimeter has been used to measure simultaneously the specific heat, thermal conductivity and thermal diffusivity around the antiferromagnetic to paramagnetic phase transition in $KMnF_3$. It has been found that the critical exponent and the amplitude ratio of the thermal diffusivity are the same as those of the specific heat. Both values agree with the predictions of the three-dimensional Heisenberg model for isotropic magnets, although this transition is not only magnetic but structural as well. The thermal conductivity shows a broad peak at the Néel temperature that could be related to a reduction of the phonon scattering by the spin fluctuation as the transition temperature is approached.

## I. INTRODUCTION

The critical phenomena associated to static quantities, such as specific heat ($c_p$), at magnetic phase transitions has been widely studied in the past. Most experimental results confirm the universality hypothesis which states that the critical behaviour depends only on the dimensionality of the system ($d$) and on the degree of freedom of the order parameter ($n$). For instance, antiferromagnetic materials, provided that only short-range interactions are relevant, can be described by a Heisenberg model if d = 3 and $n$ = 3 (isotropic), by an XY-model if d = 3 and $n$ = 2 and by an Ising model if d = 3 and $n$ = 1 (uniaxial anisotropic). On the contrary, in the case of dynamics quantities, like the thermal conductivity ($K$) and the thermal diffusivity ($D$), few high-resolution data for solid samples are available in the literature.[1-6] This is due to the conflict between the need of producing thermal gradients for thermal transport measurements and the need of keeping them as small as possible for not destroying the critical information. In this way the ac photopyroelectric calorimetry[7,8] is specially suited for the measurement of the thermal transport properties around phase transitions, since small temperature gradients in the sample produce a good signal-to-noise ratio, letting thermal properties be measured with high accuracy. Measurements on the uniaxial antiferromagnet $FeF_2$ show that the critical exponents and amplitude ratios of thermal diffusivity ($b$, $U^+/U^-$) and specific heat ($\alpha$, $A^+/A^-$) satisfy the relations:[3] $b = -\alpha =$ -0.11 and $U^+/U^- = A^-/A^+$ = 2.0, in agreement with the prediction of the so-called model $C$, for which the energy is conserved.[9,10] However, the thermal conductivity shows a broad peak in the vicinity of the transition temperature but no clear anomaly was detected. Measurements on the isotropic antiferromagnet $RbMnF_3$, a well known isotropic antiferromagnet, gave:[5] $b = \alpha =$ -0.11 and $U^+/U^- = A^+/A^-$ = 1.27, in agreement with recent calculations by Pawlak.[10] In thermal conductivity a small anomaly was found with a cusplike maximum and no divergence. As for both models the critical exponent for thermal diffusivity is the same (b = -0.11) the important information about anisotropy lies in the amplitude ratio and in the magnitude of the regular term.

In this work we use a photopyroelectric calorimeter to measure simultaneously $c_p$, $K$ and $D$ in $KMnF_3$. At room temperature it has a perovskite cubic structure but unlike $RbMnF_3$ it undergoes several structural phase transitions as the temperature decreases. In particular at 88 K a change from tetragonal to orthorhombic structure is superimposed to the antiferromagnetic to paramagnetic transition. Our aim is to study the critical behaviour of the thermophysical parameters near the magnetic transition as well as the influence of the nearby structural transition on these parameters. Actually, previous $c_p$ measurements are not conclusive since the obtained critical exponents are rather far from those predicted in the 3D-Heisenberg model: $\alpha$ = - 0.16 and $\alpha$ = -0.19.[11,12] Moreover, as far as we know no high-resolution measurements of $D$ and $K$ close to the magnetic transition have been performed. It is worth noting that the possibility of a simultaneous measurement of both static and dynamic quantities on the same sample under the same experimental conditions, such as thermal gradients, heating rate, etc. can greatly help when using the results of the measurements in scaling laws. Our results indicate that the critical behaviour of $c_p$ and $D$ agrees with the 3D-Heisenberg model and are not affected by the presence of the structural phase transition. Data on thermal conductivity showing a broad peak are also reported.

## II. SAMPLE CHARACTERISTICS

The density at room temperature is 3.42 g/cm$^3$. Pure $KMnF_3$ has a perovskite structure and undergoes three structural phase transitions related to the rotation of the octahedra at the following temperatures: 186.5 K, 88 K and 82 K.[13] At high temperature $KMnF_3$ shows a cubic phase with space group (*Pm3m*). Below 186.5 K it becomes tetragonal (*I4/mcm*) with the tetragonal $c$ axis developing around the axis of rotation of the octahedra ($c/a > 1$), namely, one of the [001] cubic axes. At 88 K it changes to an orthorhombic phase (*Pnma*) in which the octahedra are still rotated around the previous tetragonal axis and slightly rotated around one of the tetragonal $a$ axes. Finally below 82 K a tetragonal phase is developed in which the octahedra are rotated around all the [001] directions of the cubic phase.

Regarding its magnetic properties $KMnF_3$ becomes antiferromagnetic below the Néel temperature $T_N$ = 88 K and transform further to a canted antiferromagnet at 82 K. According to various authors an additional structural phase transition appears at 91 K, very close to $T_N$, claimed either first-order[14] or second-order.[15] Kapusta et al.[16] confirm the presence of the transition at 91 K from tetragonal to monoclinic ($P2_1/m$) instead of the transition at 83 K. However, no definitive consensus has been obtained yet.

Measurements of the uniaxial anisotropy (the ratio between the anisotropic field $H_A$ and the exchange field $H_E$) have shown different results:[17,18] $H_A/H_E = 4.3\times10^{-3}$ and $H_A/H_E = 5.9\times10^{-6}$. According to the first value $KMnF_3$ is an anisotropic antiferromagnet like $MnF_2$, while according to the second one it is an isotropic antiferromagnet, as was the case of other magnetic trifluorides like $RbMnF_3$ and $KNiF_3$.

Specific heat ($c_p$) measurements around the magnetic phase transition show the typical lambda shape of a second order phase transition.[11,12,19] Only in the work of Khlyustov et al.[19] a singularity a 95 K, different from the magnetic one, is reported. Analysis of the critical behaviour shows the following values for the critical exponent and the amplitude ratio: $\alpha$ = - 0.16 and $A^+/A^-$ = 1.73;[11] and $\alpha$ = -0.19 and $A^+/A^-$ = 1.36.[12] Although these values point to a 3D Heisenberg universality ($\alpha$ = -0.11 and $A^+/A^-$ = 1.58), a crossover to 3D Ising universality is also found as the temperature approaches $T_N$.

Thermal conductivity ($K$) measurements show a broad minimum near the Néel temperature which was firstly attributed to magnetic interactions.[20,21] However, Gustafson and Walter[22] have reported that the similar antiferromagnet $RbMnF_3$, which has no structural phase transitions, shows no anomalies in the thermal conductivity near $T_N$. Since $RbCaF_3$, which has similar structural transitions to those observed in $KMnF_3$ but is nonmagnetic, shows a similar minimum in thermal conductivity it was concluded that the reduction in the thermal conductivity is due to the enhanced phonon-phonon scattering due to the soft phonon modes and not to the spin-phonon coupling.[23] However, probably the poor resolution in the thermal conductivity measurements is the reason why no magnetic anomaly has been measured previously. Accordingly a high resolution technique is needed to detect a variation in $K$ in the vicinity of a magnetic transition in order to study its critical behaviour.

## III. EXPERIMENTAL SETUP AND RESULTS

Simultaneous measurements of $K$, $D$ and $c_p$ have been performed by a high-resolution ac photopyroelectric calorimeter used in the standard back-detection configuration.[7,8] An acousto-optically modulated He-Ne laser beam of 5 mW illuminates the front surface of the sample under study. Its rear surface is in thermal contact with a 350 μm thick $LiTaO_3$ pyroelectric detector with Ni-Cr electrodes on both surfaces, by using an extremely thin layer of a highly heat-conductive silicone grease (Dow Corning, 340 Heat Sink Compound). The photopyroelectric signal is processed by a lock-in amplifier operating in the current mode. Both sample and detector are placed inside a nitrogen bath cryostat that allows measurements in the temperature range from 77 K to 500 K, at rates that vary from 100 mK/min for measurements on a wide temperature range to 10 mK/min for high-resolution runs close to the phase transitions.

The natural logarithm ($\ln V$) and the phase ($\Psi$) of the normalized photopyroelectric current (that is obtained dividing by the signal provided by the bare detector) have a linear dependence on $\sqrt{f}$, with the same slope. From their slope $m$ and from the vertical separation $d$, the thermal diffusivity and the thermal effusivity ( $e = \sqrt{\rho c_p K} = K/\sqrt{D}$ ) of the sample can be obtained[24]

$$D = \frac{\pi \ell^2}{m^2}, \tag{1}$$

$$e = e_p \left( \frac{2}{\exp(d)} - 1 \right), \tag{2}$$

where $\rho$ is the density, $\ell$ is the sample thickness and $e_p$ is the thermal effusivity of the pyroelectric detector. These equations are valid for opaque and thermally thick samples (i.e. $\ell$ is higher than the thermal diffusion length $\mu = \sqrt{D/\pi f}$). However, at temperatures below 230K, where the coupling grease freezes, an important piezoelectric contribution is added to the pyroelectric signal, affecting the linearity of the dependence on $\sqrt{f}$. This contribution has been taken into account in order to obtain reliable values of $D$ and $e$.[25]

Once thermal diffusivity and effusivity have been measured at a certain reference temperature ($T_{ref}$, $D_{ref}$, $e_{ref}$), the temperature is changed while recording the amplitude and phase of the pyroelectric signal, at a fixed frequency. The temperature dependence of $D$ and $e$ are given by:[24,26]

$$D(T) = \left( \frac{1}{\sqrt{D_{ref}}} - \frac{\Delta(T)}{\ell \sqrt{\pi f}} \right)^{-2} \tag{3}$$

$$e(T) = e_p(T) \left( \frac{1 + \frac{e_{ref}}{e_p(T_{ref})}}{\exp[\Delta''(T)]} - 1 \right), \tag{4}$$

where $\Delta(T) = \Psi(T) - \Psi(T_{ref})$, $\Delta'(T) = \ln V(T) - \ln V(T_{ref})$ and $\Delta''(T) = \Delta'(T) - \Delta(T)$. Finally, the temperature dependence of specific heat and thermal conductivity are calculated from the following relations:

$$c_p(T) = \frac{e(T)}{\rho \sqrt{D(T)}} \tag{5}$$

$$K(T) = e(T) \sqrt{D(T)} \tag{6}$$

This technique is specially suited for the measurement of thermal properties around phase transitions, since small temperature gradients in the sample produce a good signal-to-noise ratio, letting thermal parameters be measured with high accuracy.

We have measured a single crystal plate of $KMnF_3$ whose thickness is 523 μm. Since it is transparent to the He-Ne laser, the front surface was coated with a very thin metallic layer. Its thermal properties have been measured in the temperature range from 78 K to 300 K. A modulation frequency of 8.0 Hz, high enough to assure the sample is thermally thick but low enough to guarantee a good signal-to-noise ratio, has been used. Three anomalies in the thermal properties associated to phase transitions have been observed at 186.2 K, 86.6 K and 80.5 K. No anomaly at 91 K has been found. High-resolution measurements around each of the transition temperatures have been performed.

In fig. 1 the temperature dependence of $K$, $D$ and $c_p$ around the structural phase transition from cubic to tetragonal is shown. As can be seen a narrow and sharp dip in thermal diffusivity together with a similarly narrow and abrupt peak in specific heat are obtained. Both show a small thermal hysteresis of about 0.2 K. This result confirms the first order nature of this transition. On the other hand, thermal conductivity presents a small jump with no anomaly. The temperature dependence of $D$ around the two low temperature phase transitions is shown in fig. 2. A dip with an inverted lambda shape and no hysteresis, characteristic of a second order phase transition, has been found at the Néel temperature $T_N = 86.6$ K. On the other hand, an abrupt jump in diffusivity appears at 81 K showing a large hysteresis of about 1.5 K. Our thermal measurements do not show any transition at 91 K, although it has been claimed by several authors.

## IV. FITTING PROCEDURE AND DISCUSSION

In this section we concentrate on the study of the critical behaviour of the magnetic transition at 86.6 K. Experimental measurements of $c_p$, $D$ and $1/K$ in the vicinity of $T_N$ are shown by dots in fig. 3. As can be seen the $D$ values are very accurate since they are obtained directly form the phase of the photopyroelectric signal (see Eq. (3)), while the $c_p$ and $1/K$ values are affected by a bit higher uncertainty since they are obtained by combining both, amplitude and phase (see Eqs. (5) and (6)). Note that the scale for thermal conductivity has been chosen so as to show that its singularity is much smaller than those obtained for $D$ and $c_p$.

The critical behaviour of the specific heat is described by a function of the form

$$c_p = B + Ct + A^{\pm}|t|^{-\alpha}\left(1 + E^{\pm}|t|^{0.5}\right), \qquad (7)$$

where $t = (T - T_N)/T_N$ is the reduced temperature, and α, $A^{\pm}$, $B$, $C$ and $E^{\pm}$ are adjustable parameters. Superindexes + and - stand for $T > T_N$ and $T < T_N$ respectively. The linear term represents the background contribution to the specific heat, while the last term is the magnetic contribution to the specific heat. The factor under parenthesis is the correction to scaling that represents a singular contribution to the leading power as known from experiments and theory. [27,28]

For thermal diffusivity and thermal conductivity we assume, similarly to the specific heat, that they are given by a sum of a regular term plus a singular one, with a correction to scale term[1,5]

$$D = V + Wt + U^{\pm}|t|^{-b}\left(1 + F^{\pm}|t|^{0.5}\right) \tag{8}$$

$$1/K = 1/K_{nonmag} + 1/K_{mag} = L + Mt + N^{\pm}|t|^{-g}\left(1 + H^{\pm}|t|^{0.5}\right). \tag{9}$$

Subindex "nonmag" refers to phonon-phonon scattering, umklapp, scattering with impurities, etc., while "mag" refers to the spin-phonon scattering mechanisms. The magnetic resistive term originates from the spin-lattice interaction and close to $T_N$ from an additional term that accounts for phonon scattering by critical fluctuations of the order parameter. The reason for fitting the $1/K$ data instead of the $K$ ones relies on the fact that all the thermal resistances associated to the various heat conduction mechanisms in the sample are in series. In this way the singular term in Eq. (9) can be related to the magnetic contribution to the heat conduction processes.

The data were simultaneously fitted for $T > T_N$ and $T < T_N$ with a non-linear least square routine. First of all, we selected a fitting range close to the peak (dip) while avoiding the rounding, and kept fixed the value of $T_N$. We obtained a first fitting without the correction to scaling term and obtained a set of adjusted parameters. Afterwards, we tried to increase the number of points included in the fitting, first fixing $t_{min}$ and increasing $t_{max}$, and then fixing $t_{max}$ and decreasing $t_{min}$. The next step was introducing the correction to scaling term trying to improve the fitting. As a last checking, we let $T_N$ be a free parameter in order to confirm the fitting. In the whole process, we focused our attention on the rms deviations as well as on the deviation plots, which are the plots of the difference between the fitted values and the measured ones as a function of the reduced temperature. The continuous lines in fig. 3 represent the fittings of the experimental data to Eqs. (7), (8) and (9). In fig. 4 the deviation plots show the range of the reduced temperatures used in the fittings as well as their quality. The values of the fitting parameters are given in table 1.

## A. Specific heat

It is seen that the values of the critical exponent and the amplitude ratio of the specific heat are in excellent agreement with the predictions of the 3D-Heisenberg model for isotropic magnets ($\alpha$ = -0.11 and $A^+/A^-$ = 1.58).[3,29] Anyway, the presence of a small anisotropy can produce a crossover from the 3D-Heisenberg universality class for isotropic magnets to the 3D-Ising model for uniaxial anisotropic magnets as the reduced temperature approaches zero. Following the renormalization group theory[31] a crossover temperature $t_x$, related to the degree of anisotropy of the material, can be defined: $|t_x| = |H_A / H_E|^{0.8}$. Accordingly, for $|t| > |t_x|$ the system is expected to exhibit a 3D-Heisenberg universality class while for $|t| < |t_x|$ the system is expected to behave rather like a 3D-Ising model. The values of $|t_x|$ for several antiferromagnets are given in fig. 5.[12,17,18,32-36] For instance, $FeF_2$ and $CoF_2$, two well known uniaxial antiferromagnets, agree with the 3D-Ising model in the whole region of reduced temperature. On the other hand, $RbMnF_3$, an acknowledged isotropic antiferromagnet, agrees with the 3D-Heisenberg model, since crossover to 3D-Ising model is expected only at $|t| < 6\times10^{-5}$, which has been never reached experimentally. An intermediate value appears for weakly anisotropic $Cr_2O_3$, for which evidence of crossover has been found as the range of the reduced temperature used in the fittings goes below $10^{-3}$.[1,37] As can be seen in fig. 5 two different values for $|t_x|$ are available in the literature for $KMnF_3$. According to the highest one a crossover to 3D-Ising behaviour should be expected and it was used by Akutsu et al.[12] to explain the deviation of its critical exponent value (α = -0.19) from the pure 3D-Heisenberg prediction. Our results support, in a wide temperature range ($7\times10^{-5}$ - $3\times10^{-2}$), the Heisenberg model behaviour and therefore we conclude that only the lowest $t_x$ value, corresponding to $H_A/H_E = 5.9\times10^{-6}$,[18] is appropriate for $KMnF_3$ which means an extremely weak anisotropy.

### B. Thermal diffusivity

The critical exponent and the amplitude ratio of the thermal diffusivity are very similar to those found for specific heat, a result that was already found for $RbMnF_3$.[5] Both agree with the recent predictions by Pawlak[10] about the critical thermal diffusivity of isotropic magnets, following the phenomenological hydrodynamical approach to critical dynamics: $b = \alpha$ and $U^+/U^- = A^+/A^-$. It is worth noting that the same prediction is also found if the thermal conductivity shows no singularity at $T_N$ or if it is very small. In such a case, using Eq. (7) for the critical behaviour of $c_p$ the thermal diffusivity can be written as

$$D = \frac{K}{\rho c_p} = \frac{K}{\rho B}\left(1 + \frac{E}{B}t + \frac{A^{\pm}}{B}|t|^{-\alpha}\right)^{-1}, \qquad (10)$$

where we have neglected the correction to scale factor in Eq. (7), which is introduced to increase the fitting temperature range. Two universality classes will be considered:

(a) 3D-Heisenberg universality ($A^{\pm} < 0$, $A^+/A^- = 1.58$, $\alpha = -0.11$). In this case as the reduced temperature approaches 0, $|t|^{-\alpha} << 1$ and therefore Eq. (10) can be expanded as

$$D = \frac{K}{\rho B}\left(1 - \frac{E}{B}t - \frac{A^{\pm}}{B}|t|^{-\alpha}\right) = V + Wt + U^{\pm}|t|^{-b}. \tag{11}$$

According to this last expression the following relations for the universal parameters are found: $b = \alpha = -0.11$, $U^{\pm} = -KA^{\pm}/\rho B^2 > 0$ and $U^+/U^- = A^+/A^- = 1.58$. Similarly, for the non-universal parameters it is found: $V = K/\rho B$ and $W = -KE/\rho B^2$. Using the results of table 1 we obtain for $KMnF_3$ a good correlation between the fitted values of the thermal diffusivity parameters and those obtained from the above relations: $b = -0.123 \approx \alpha = -0.111$, $U^- = 0.200 \approx -KA^-/\rho B^2 = 0.215$, $U^+/U^- = 1.51 \approx A^+/A^- = 1.41$, $V = 1.09 \approx K/\rho B = 1.00$ and $W = -0.019 \approx -KE/\rho B^2 = -0.007$. Only for $W$ the discrepancy is significant, but it is due to its low value. A good correlation is also found for $RbMnF_3$ using the fitting parameters of table I in Ref. 5: $b = -0.11 = \alpha = -0.11$, $U^- = 0.011 \approx -KA^-/\rho B^2 = 0.014$, $U^+/U^- = 1.27 = A^+/A^- = 1.27$, $V = 0.067 \approx K/\rho B = 0.168$ and $W = -0.003 \approx -KE/\rho B^2 = -0.005$.

(b) 3D-Ising universality ($A^{\pm} > 0$, $A^+/A^- = 0.50$, $\alpha = +0.11$). Now as the reduced temperature approaches 0, $|t|^{-\alpha} \to \infty$ and therefore Eq. (10) reduces to

$$D = \frac{K}{\rho A^{\pm}}|t|^{\alpha} = V + Wt + U^{\pm}|t|^{-b}. \tag{12}$$

Accordingly, the following relations for the universal parameters are found: $b = -\alpha = -0.11$, $U^{\pm} = K/\rho A^{\pm} > 0$ and $U^+/U^- = A^-/A^+ = 2.00$; while for the non-universal parameters: $V, W \to 0$. Using the results for $FeF_2$ in Ref. 3 a good correlation between the fitted values of the diffusivity parameters and those obtained from the above relations: $b = -0.11 \approx -\alpha = -0.11$, $U^- = 0.114 \approx K/\rho A^- = 0.121$, $U^+/U^- = 1.97 \approx A^-/A^+ = 2.00$, $V = 0.005 << U^- = 0.114$ and $W = 0.011 << U^- = 0.114$.

These results together confirm the predictions of Pawlak and suggest that thermal diffusivity plays a similar role in critical behaviour as the specific heat.

### C. Thermal conductivity

As can be seen in table 1, the singular term of $1/K$ has a positive amplitude ($N^+, N^- > 0$) and a negative exponent ($g < 0$), indicating that the thermal conductivity does not diverge at $T_N$. Moreover, the values of the critical exponent and the amplitude ratio ($g = -0.10$ and $N^+/N^- = 1.15$) are very similar to those found by Marinelli et al. for $RbMnF_3$ ($g = -0.08$ and $N^+/N^- = 1.1$).[5] Both results are close to those expected for the 3D-Heisenberg model for specific heat, although its meaning is not well understood.

Up to now few works showing high resolution thermal conductivity data in the vicinity of the magnetic transition temperature are available in the literature. In some of them no singular behaviour has been found, e.g. $Cr_2O_3$, CoO and EuO.[1,2,6] However, in Gd a clear dip was reported,[4] while a broad peak has been obtained for $FeF_2$, $RbMnF_3$ and in the present work for $KMnF_3$.[3,5] These results indicate that the critical behaviour of $K$ shows a system dependency that is probably due to the different spin-phonon coupling mechanisms that are present in particular compounds. Although the reasons for the broad peak in the above mentioned magnetic fluorides are not understood in detail, we speculate that a reduction of the phonon scattering by the spin fluctuation as $T_N$ is approached is playing an important role.

Existing theories on the critical behaviour of thermal conductivity are largely incomplete. Kawasaki observed that since the heat transport is dominated by short range processes, such as phonon scattering, no divergence should be expected for $K$.[38] Moreover, as these effects are system dependent, he suggested that the scaling laws may not hold for heat transport. The dynamic renormalization group theory[9,10,39] predicts that a renormalization of the bare thermal conductivity requires a contribution to the self-energy of the form $i\omega/q^2$, which is divergent as the wave vector $q \rightarrow 0$ for a fixed non-zero frequency $\omega$. But away from criticality the self-energy is finite in the $q \rightarrow 0$ limit at all orders in perturbation theory. Thus, these perturbations which were considered so far can generate no corrections to $K$, and the thermal conductivity will be given exactly by its "bare" value. These results together with the few experimental data are the reasons why in modern theories of critical phenomena the heat transport has been systematically ignored. As a consequence, even in the simple case of magnetic insulators it is not clear which critical behaviour for $K$ should be expected.

A possible reason for the small peak in $K$ for $FeF_2$, $RbMnF_3$ and $KMnF_3$ is that it could be an artefact raised by the way in which this quantity is obtained from the experimental data. From the amplitude and phase of the photopyroelectric signal thermal diffusivity and effusivity are obtained, which at the magnetic transition show a dip and a peak respectively. Then thermal conductivity is obtained from Eq. (6). This means that close to the magnetic transition $K$ is the result of multiplying a peak by a dip that could produce artificial peaks or dips. However, we have measured the thermal conductivity of other magnetic materials that do not show singularity but only a small change in slope, as is the case of CoO, MnO, $Cr_2O_3$, $Rb_2MnF_4$, $KFeF_4$, among others.[40] Therefore we conclude that the peak is real although it is much weaker than the peak in specific heat or the dip in thermal diffusivity.

Another possible explanation of the anomalous behaviour of $K$ could be the volume magnetostriction. Let us recall a simple kinetic expression for thermal conductivity

$$K = \frac{1}{3}\sum_{q,\lambda} v_{q,\lambda} l_{q,\lambda} \frac{\partial u_{q,\lambda}}{\partial T}, \qquad (13)$$

where $v_{q,\lambda}$, $l_{q,\lambda}$, $u_{q,\lambda}$ denote velocity, mean free path and energy of the phonons with wave vector $q$. The phonon energy density $u_{q,\lambda} = \frac{\hbar\omega_{q,\lambda}}{V(e^{\hbar\omega_{q,\lambda}/k_B T} - 1)}$ depends on temperature explicitly through the factor $k_B T$ and also through the volume $V$. This last one is temperature dependent due to the magnetostrictive coupling, through which there is a contribution to the volume proportional to $\langle \Phi^2 \rangle$, where $\Phi$ is the order parameter.[10] It induces anomalous contribution to $\frac{\partial u_{q,\lambda}}{\partial T}$:

$$K = \frac{1}{3}\sum_{q,\lambda} v_{q,\lambda} l_{q,\lambda} \left[ \left( \frac{\partial u_{q,\lambda}}{\partial T} \right)_{regul} + \left( \frac{\partial u_{q,\lambda}}{\partial T} \right)_{anom} \right] = \frac{1}{3}\sum_{q,\lambda} v_{q,\lambda} l_{q,\lambda} \left[ \left( \frac{\partial u_{q,\lambda}}{\partial T} \right)_{regul} + \left( \frac{\partial u_{q,\lambda}}{\partial V} \frac{\partial V}{\partial T} \right) \right] =$$

$$= \frac{1}{6\pi^2} \sum_{\lambda} \int d^3\vec{q} v_{q,\lambda} l_{q,\lambda} \frac{(\hbar\omega_{q,\lambda})^2 e^{\hbar\omega_{q,\lambda}/k_B T}}{k_B T^2 \left( e^{\hbar\omega_{q,\lambda}/k_B T} + 1 \right)^2} \left[ 1 + \gamma_{q,\lambda} \left( 1 - f(\hbar\omega_{q,\lambda}/k_B T) \right) \frac{T}{V} \frac{\partial V}{\partial T} \right] =$$

$$= K_0 \left[ 1 + ag\gamma^{eff} T \frac{\partial \langle \Phi^2 \rangle}{\partial T} \right], \qquad (14)$$

where $\gamma_{q,\lambda} = -\frac{V}{\omega_{q,\lambda}} \frac{\partial \omega_{q,\lambda}}{\partial V}$, $f(\hbar\omega_{q,\lambda}/k_B T) = \frac{k_B T}{\hbar\omega_{q,\lambda}} \frac{e^{\hbar\omega_{q,\lambda}/k_B T} - 1}{e^{\hbar\omega_{q,\lambda}/k_B T}}$ and $\frac{\partial \langle \Phi^2 \rangle}{\partial T} \propto \langle \Phi^2 \Phi^2 \rangle$ is proportional to the magnetic specific heat $c_m$. To sum up, the thermal conductivity has an anomalous contribution proportional to the magnetic specific heat, to the magnetostrictive coupling constant $g'$ and to a complicated integral from Grüneisen parameters ($\gamma^{eff}$): $K = K_0 + ag'\gamma^{eff} T c_m$, where $a$ is a constant. The importance of the anomalous contribution to $K$ depends on the thermal expansion through $g'$. To the best of our knowledge, no thermal expansion data near the magnetic transition in $KMnF_3$ are known. However, the dynamic elastic response of $KMn_{1-x}Ca_xF_3$ ($x < 0.018$) has been recently measured by Dynamic Mechanical Analysis (DMA7-Perkin Elmer)[41] and continuous wave resonance (cw-) technique around the three phase transitions.[42] A negative dip anomaly at $T_N$ in the elastic constant $C_{11}$ of about 25%, which indicates a considerable magnitude of the magnetostrictive coupling coefficient $g'$, has been found.

## V. CONCLUSIONS

High resolution measurements of $c_p$, $D$ and $K$ around the antiferromagnetic to paramagnetic transition of $KMnF_3$ have been performed using a photopyroelectric calorimeter. It has been found for the first time that the critical behaviour of $c_p$ agrees unambiguously with the 3D-Heisenberg universality class. It means that the presence of a simultaneous phase transition from orthorhombic to tetragonal at the magnetic transition temperature does not affect the critical exponents. On the other hand, the critical exponent and the amplitude ratio of $D$ are the same as those found for $c_p$, confirming the recently published theoretical predictions.[10] In its turn, $K$ shows a broad peak close to the magnetic transition probably related to a reduction of the phonon scattering by the spin fluctuation in the vicinity of $T_N$. Finally, we expect that these high resolution measurements of $K$ together with those performed in $FeF_2$ and $RbMnF_3$ stimulate theoretical studies on the critical behaviour of this quantity in order to fill the existing void.

## ACKNOWLEDGMENTS

This work has been supported by the University of the Basque Country through research grant No. EHU06/24 and by the Austrian FWF project No. P19284-N20.

## FIGURE CAPTIONS

FIG. 1. Temperature dependence of $K$, $D$ and $c_p$ around the structural transition from cubic to tetragonal.

FIG. 2. Temperature dependence of $D$ in the vicinity of the two low temperature phase transitions.

FIG. 3. Temperature dependence of $c_p$, $D$ and $1/K$ in the vicinity of $T_N$. Dots are the experimental results and the continuous lines correspond to the best fits.

FIG. 4. Deviation plots corresponding to the fits of figure 3. Open circles are for $T < T_N$ and crosses for $T > T_N$.

FIG. 5. Values of $|t_x|$ for several antiferromagnets.

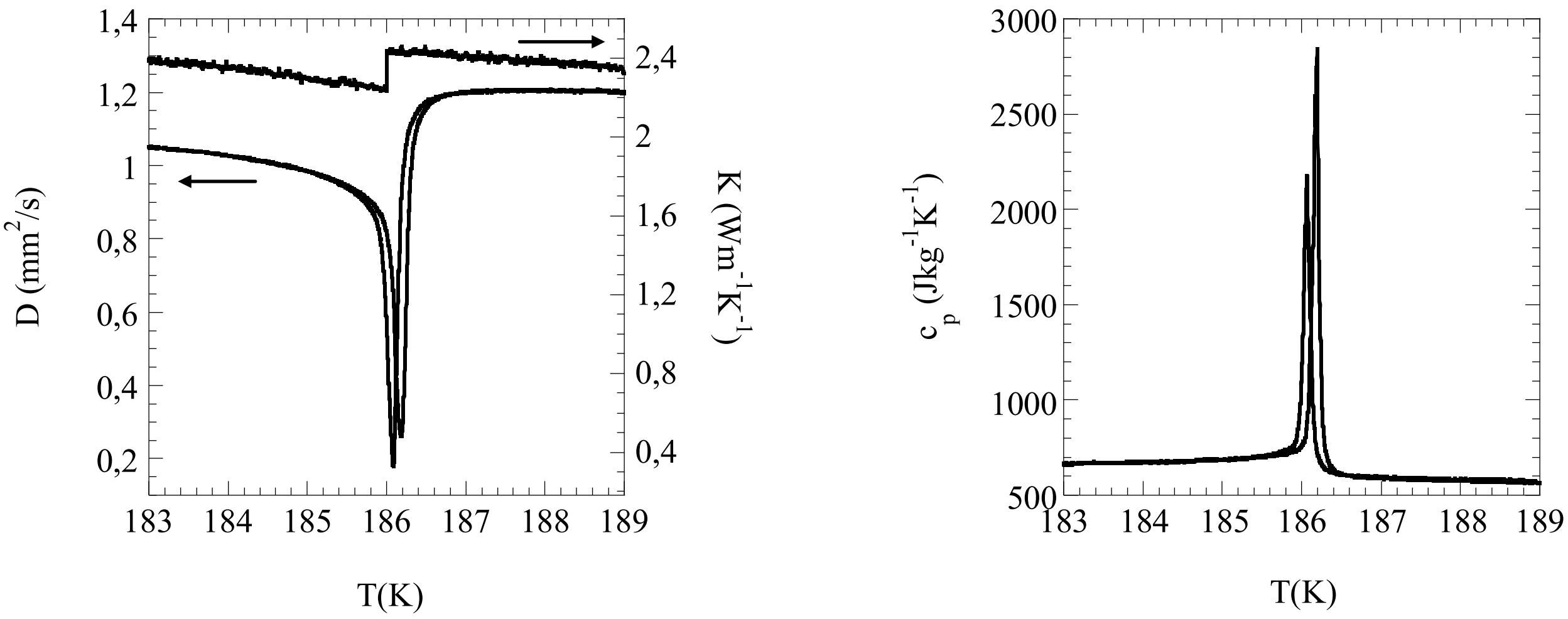

FIG. 1. A. Salazar et al.

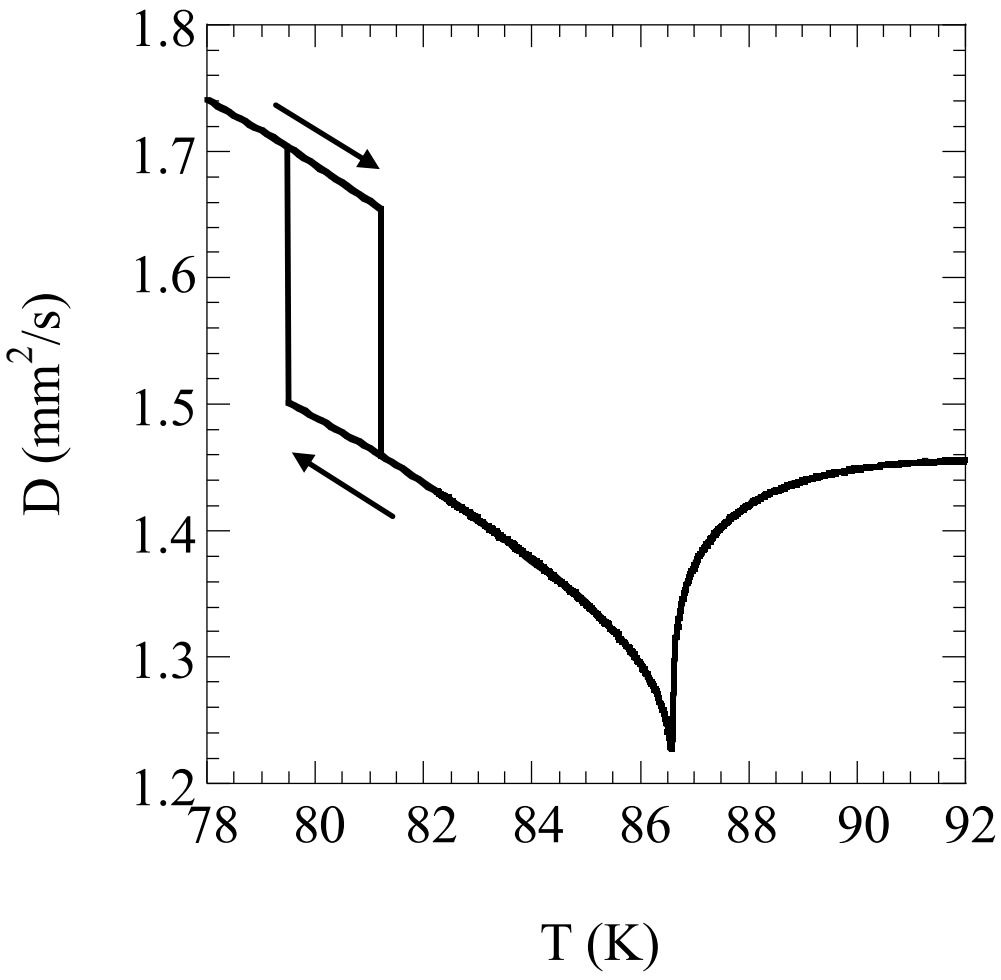

FIG. 2. A. Salazar et al.

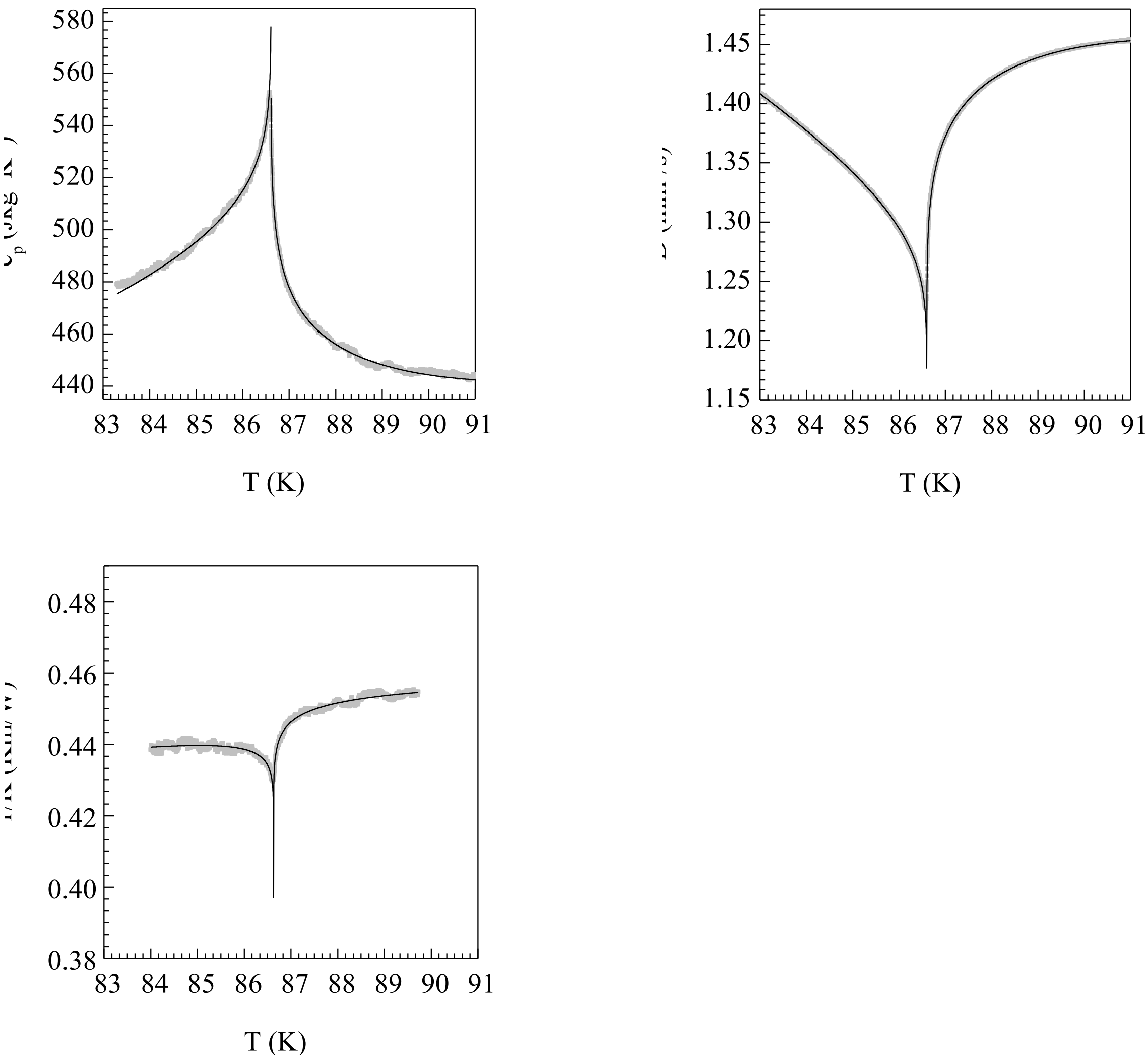

FIG. 3. A. Salazar et al.

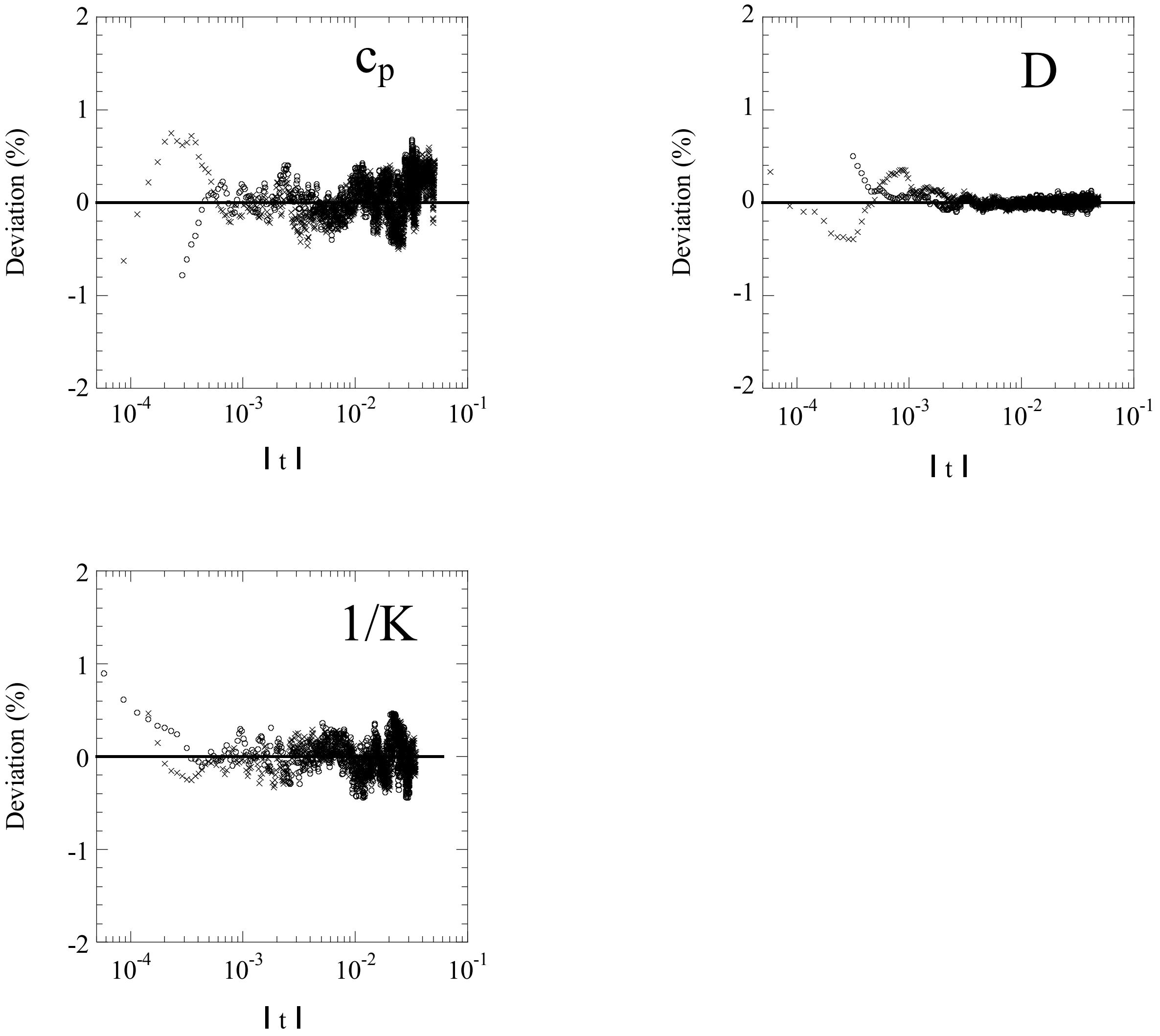

FIG. 4. A. Salazar et al.

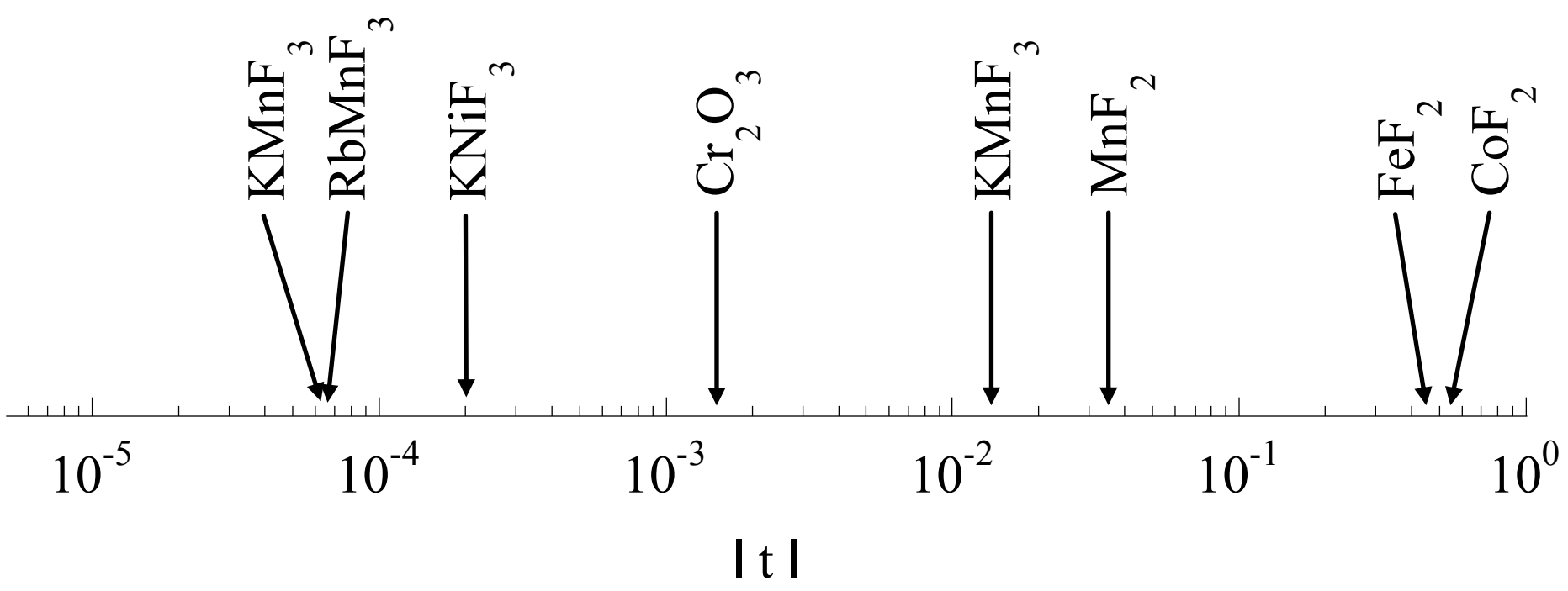

FIG. 5. A. Salazar et al.

TABLE I. Values of the adjustable fitting parameters for specific heat, thermal diffusivity and the inverse of thermal conductivity obtained with Eqs. (7), (8) and (9) respectively. In bold are the universal parameters.

| | $\alpha$ | $A^+/A^-$ | $T_N$ (K) | $B$ ($Jkg^{-1}K^{-1}$) | $E$ ($Jkg^{-1}K^{-1}$) | $A^-$ ($Jkg^{-1}K^{-1}$) | $D^-$ | $D^+$ | $R$ |
|---|---|---|---|---|---|---|---|---|---|
| $c_p$ ($Jkg^{-1}K^{-1}$) | **-0.111 ±0.002** | **1.41 ±0.05** | 86.61 ±0.01 | 650 ±3 | 4.54 ±0.07 | -140 ±3 | 0.003 ±0.001 | 0.004 ±0.002 | 0.9985 |

| | $b$ | $U^+/U^-$ | $T_N$ (K) | $V$ ($mm^2/s$) | $W$ ($mm^2/s$) | $U^-$ ($mm^2/s$) | $F^-$ | $F^+$ | $R$ |
|---|---|---|---|---|---|---|---|---|---|
| $D$ ($mm^2/s$) | **-0.123 ±0.006** | **1.51 ±0.09** | 86.6 0 ±0.01 | 1.09 ±0.01 | -0.019 ±0.001 | 0.201 ±0.008 | 0.032 ±0.004 | 0.113 ±0.007 | 0.9998 |

| | $g$ | $N^+/N^-$ | $T_N$ (K) | $L$ (Km/W) | $M$ (Km/W) | $N^-$ (Km/W) | $H^-$ | $H^+$ | $R$ |
|---|---|---|---|---|---|---|---|---|---|
| $1/K$ (Km/W) | **-0.10 ±0.03** | **1.15 ±0.35** | 86.62 ±0.01 | 0.39 ±0.02 | 0.0006 ±0.0002 | 0.056 ±0.009 | -0.10 ±0.02 | -0.07 ±0.03 | 0.9803 |